\documentclass[%
 preprint,
superscriptaddress,
 amsmath,amssymb,
 aps,
 prl,
]{revtex4-2}

\usepackage[pdftex]{graphicx}
\usepackage{color,latexsym,inputenc,setspace,multirow,array,tabularx,bm}
\usepackage{amsfonts,amssymb,amsmath,braket}
\usepackage[colorlinks=true]{hyperref}
\usepackage{comment}

\newcommand{\bacovo}{BaCo$_{2}$V$_{2}$O$_{8}$}

\newcommand{\bacovomg}{Ba(Co$_{1-x}$Mg$_x$)$_2$V$_{2}$O$_{8}$}

\begin{document}

\title{Non-magnetic impurities as a quantum microscope of \\ finite size effects in spin chains}

\author{Quentin Faure}
\email[Corresponding author: ]{quentin.faure@cea.fr}
\affiliation{Laboratoire L\'eon Brillouin, CEA, CNRS, Universit\'e Paris-Saclay, CEA-Saclay, 91191 Gif-sur-Yvette, France}

\author{Shintaro Takayoshi}
\affiliation{Department of Physics, Konan University, 658-8501 Kobe, Japan}

\author{Virginie Simonet}
\affiliation{Universit\'e Grenoble Alpes, CNRS, Institut N\'eel, 38042 Grenoble, France}

\author{B\'{e}atrice Grenier}
\affiliation{Universit\'e Grenoble Alpes, CEA, IRIG, MEM, MDN, 38042 Grenoble, France}

\author{Stéphane Raymond}
\affiliation{Universit\'e Grenoble Alpes, CEA, IRIG, MEM, MDN, 38042 Grenoble, France}

\author{Eric Ressouche}
\affiliation{Universit\'e Grenoble Alpes, CEA, IRIG, MEM, MDN, 38042 Grenoble, France}

\author{Xavier Fabrèges}
\affiliation{Laboratoire L\'eon Brillouin, CEA, CNRS, Universit\'e Paris-Saclay, CEA-Saclay, 91191 Gif-sur-Yvette, France}

\author{Pascal Lejay}
\affiliation{Universit\'e Grenoble Alpes, CNRS, Institut N\'eel, 38042 Grenoble, France}

\author{Thierry Giamarchi}
\affiliation{Department of Quantum Matter Physics, University of Geneva, 1211 Geneva, Switzerland}

\author{Sylvain Petit}
\affiliation{Laboratoire L\'eon Brillouin, CEA, CNRS, Universit\'e Paris-Saclay, CEA-Saclay, 91191 Gif-sur-Yvette, France}

\date{\today}

\begin{abstract}

\textbf{Finite-size effects are ubiquitous in quantum matter, yet typically invisible in the thermodynamic limit; when system size is restricted, boundaries and confinement reveal modified spectra, correlations, and emergent phenomena. With growing access to well-controlled few-spin systems, these effects are now directly observable and offer a route to probing universal physics and enabling quantum information applications. Here, we study finite-size physics in the quasi-one-dimensional Ising-like quantum antiferromagnet Ba(Co$_{1-x}$Mg$_x$)$_2$V$_2$O$_8$ with $x = 2\%$ and $x = 5\%$, where non-magnetic Mg$^{2+}$ ions fragment the Co$^{2+}$ spin chains into finite segments. Combining inelastic neutron scattering and numerical calculations, we identify distinct regimes with fundamentally different low-energy quantum excitations: spinons confined to chain boundaries, and a cluster mode in shorter chains corresponding to a collective reversal of the N\'eel-ordered segment without classical analogue. Our results demonstrate how chemical substitution acts as a quantum microscope providing direct access to boundary-shaped excitations in strongly correlated spin systems.}

\end{abstract}

\maketitle

\section*{Introduction}

Quantum spin chains constitute a central platform for exploring strong correlations and emergent quasiparticles in one dimension. Their reduced dimensionality enhances quantum fluctuations, suppresses classical long-range order, and stabilizes excitation continua arising from fractionalized spinons---mobile domain walls whose neutron-scattering signatures differ fundamentally from conventional magnons~\cite{Giamarchi2003, Mourigal2013, Lake2013}. Although most theoretical insights originate from the thermodynamic limit, growing attention has shifted toward finite chains, where boundaries, discrete spectra, and length-dependent coherence give rise to qualitatively new physics.

Beyond their role in model magnetism, finite spin chains are also relevant for quantum information science. Entire segments can encode and transmit quantum information through collective degrees of freedom, enabling proposals for coherent state transfer without single-site control~\cite{Bose2007, Meier2003PRL, Meier2003PRB, Bose2005, Christandl2004, Kay2006}. The transverse-field Ising model (TFIM), a cornerstone of quantum many-body physics~\cite{Pfeuty1970, Sachdev_book}, provides a paradigmatic example: via the Jordan–Wigner transformation, it maps onto free fermions, allowing an exact description of quantum criticality and serving as a benchmark for quantum simulation. This has motivated high-fidelity experimental realizations in Rydberg arrays, trapped ions, and cold atoms~\cite{Simon2011, Bernien2017, Schauss_2018, Monroe2021, King2022, Mi2024}, where finite size is not a correction but a defining feature governing entanglement and low-energy structure.

The Jordan–Wigner transformation also provides a natural interface with 1D topological models such as the Su-Schrieffer-Heeger model, which first predicted topological edge states and the Kitaev chain, where these boundary modes become Majorana zero modes~\cite{SSH1979, Kitaev2001_chain, ReadGreen2000}.
In particular, the TFIM and $XY$ models map to quadratic fermions with pairing terms, linking magnetic boundary modes to Majorana-like physics~\cite{Fendley_2012,  Fendley_2016}. Away from these limits, additional interactions, as in $XXZ$ systems for instance, move the model beyond the free-fermion regime, but boundary physics can persist in the form of edge modes, gapped excitations, and parity-dependent splittings~\cite{Kattel2025_2, Kattel2025, Pasnoori2025}. Spin chains therefore offer a versatile setting to explore boundary-localized excitations beyond idealized one-dimensional topological models.

Controlled realizations of finite chains are now achievable in the solid state; atomically precise nanographene synthesis enables chains with tunable length and edge termination~\cite{Mishra2021, Zhao2025}. A highly flexible alternative is provided by nonmagnetic doping in bulk quantum magnets, which segments long chains into finite pieces. Owing to their one-dimensional nature, such systems are remarkably sensitive to intrachain disorder~\cite{Alloul2009}. Theory predicts even-odd effects~\cite{Eggert1992}, impurity-pinned edge modes~\cite{Wessel2000, Haque2010}, and boundary-induced Kondo phenomena~\cite{Laflorencie2008}, while experimentally, impurity segmentation produces notable modifications of ground states and excitation spectra in spin-1/2 and Haldane chains~\cite{Azuma1997, Fujiwara1998, Ohsugi1999, Hase1993, Masuda1998, Grenier1998, Grenier2002, Augier1999, Kenzelmann2003},

In this study, we investigate finite-size quantum effects in the Ising-like ($XXZ$) $S=1/2$ antiferromagnetic chain compound \bacovo. By replacing a fraction of the magnetic sites with non-magnetic impurities, we segment the chains into finite pieces whose length distribution is tunable through impurity density. Combining inelastic neutron scattering in this material with numerical simulation of the corresponding model, we identify three distinct low-energy regimes governed by the segmented chain length. While the shortest-chain Anderson-tower regime requires strong dilution and is accessed theoretically, our experiments access the intermediate and long-chain regimes. In these regimes, the lowest-energy excitations take the form of collective cluster modes -- a coherent reversal of the Néel states -- at intermediate lengths, and of edge-localized spinon modes stabilized by open boundaries in longer chains. These results demonstrate how impurity engineering in bulk materials provides a powerful route to resolve finite-size quantum phenomena typically accessible only in engineered nanoscale or quantum-simulator platforms.

\section*{Results}

\subsection*{\bacovo: a quasi-one-dimensional Ising quantum antiferromagnet}

Before presenting our results, we first introduce the key properties of the pure \bacovo\ system. It consists of one-dimensional screw chains of Co$^{2+}$ ions aligned along the crystallographic $c$-axis. These Co$^{2+}$ ions carry an effective spin-$\tfrac{1}{2}$ and are coupled via antiferromagnetic (AF) exchange interactions [Fig.~\ref{Fig0}a]. Owing to a pronounced Ising-like anisotropy and weak interchain interactions, \bacovo\ stabilizes a N\'eel-ordered AF ground state below the transition temperature $T_{\mathrm{N}} = 5.5$~K, with magnetic moments aligned along the $c$-axis~\cite{He2005} [Figs.~\ref{Fig0}b--c]. Unlike conventional antiferromagnetic spin chains, which host a continuum of spinon excitations, \bacovo\ exhibits a discrete sequence of gapped excitations, so-called Zeeman ladders. They correspond to bound states of spinons confined by the effective staggered field emerging from the N\'eel order and the interchain coupling of \bacovo\ [Fig.~\ref{Fig0}f]. The Zeeman ladders comprise both transverse modes, corresponding to fluctuations perpendicular to the ordered moments and involving an odd number of spin flips, and longitudinal modes associated with fluctuations parallel to the ordered moment direction and involving an even number of spin flips [Figs.~\ref{Fig0}d-e]. We denote the transverse and longitudinal modes as $\ket{N T}$ and $\ket{N L}$, respectively. In the Ising limit, these modes are associated with the number of flipped spins $N$. The transverse mode contributes to the transverse spin dynamical susceptibility ${\cal S}_{xx}$ (equivalently ${\cal S}_{yy}$), and the longitudinal modes contributes to the longitudinal spin dynamical susceptibility ${\cal S}_{zz}$. The dynamical susceptibilities are the Fourier transforms of the corresponding spin pair correlation functions and can be measured by neutron scattering [Figs.~\ref{Fig0}d--f]. These excitations were previously observed in both \bacovo\ and its structurally analogue compound SrCo$_2$V$_2$O$_8$~\cite{Grenier2015,Bera2014,Bera2017}.

The spin model Hamiltonian of \bacovo\ consists of intrachain and interchain parts \cite{faure_topological_bacovo}. The former reads 
\begin{align}
 \mathcal{H}_{\mathrm{intra}} =
  J \sum_{j}\sum_{\mu} \left[\epsilon
   \left(S_{j,\mu}^{x}S_{j+1,\mu}^{x}+S_{j,\mu}^{y}S_{j+1,\mu}^{y}\right)
   +S_{j,\mu}^z S_{j+1,\mu}^z\right],
\label{eq:Hamilintra}
\end{align}
where $J$ is the intrachain spin coupling, $\epsilon$ the Ising-like anisotropy, and $\mathbf{S}_{j,\mu}$ a spin-1/2 operator on $j$-th site of $\mu$-th chain. The interchain part reads: 
\begin{align}
 \mathcal{H}_{\mathrm{inter}} = J'\sum_{j}\sum_{\braket{\nu,\mu}} \boldsymbol{S}_{j,\mu} \boldsymbol{S}_{j,\nu},
\label{eq:Hamilinter}
\end{align}
where $J'$ is the interchain spin coupling and $\sum_{\braket{\mu,\nu}}$ denotes the summation over the pairs of nearest neighbor chains. For \bacovo, we used the parameters $J = 5.8$~meV, $\epsilon = 0.53$, and $J' = 0.17$~meV, determined previously since they best reproduce the excitation spectra of the undoped compound~\cite{faure_topological_bacovo}. \bacovo\ is a quasi-1D system satisfying $J \gg J'$.

For the analysis, we treat the interchain coupling by the mean field theory. In the presence of the N\'eel order along the $c$ axis, Eq.~\eqref{eq:Hamilinter} is approximated as $\mathcal{H}_{\mathrm{inter}} \simeq J'\sum_{j}\sum_{\braket{\mu}} \boldsymbol{S}_{j,\mu} \braket{\boldsymbol{S}_{j,\nu}}$. Although the actual structure of microscopic interchain coupling is likely to be more complex, this modeling was shown to give an excellent description of the pure system up to very large magnetic fields~\cite{takayoshi_bacovo_phasediag, Faure2021_transverse}. It also allows to deal with a 1D problem.

\subsection*{Realization of finite-size spin chains by non-magnetic impurities}

\begin{figure*}[thb]
\centering
\includegraphics[width=0.95\linewidth]{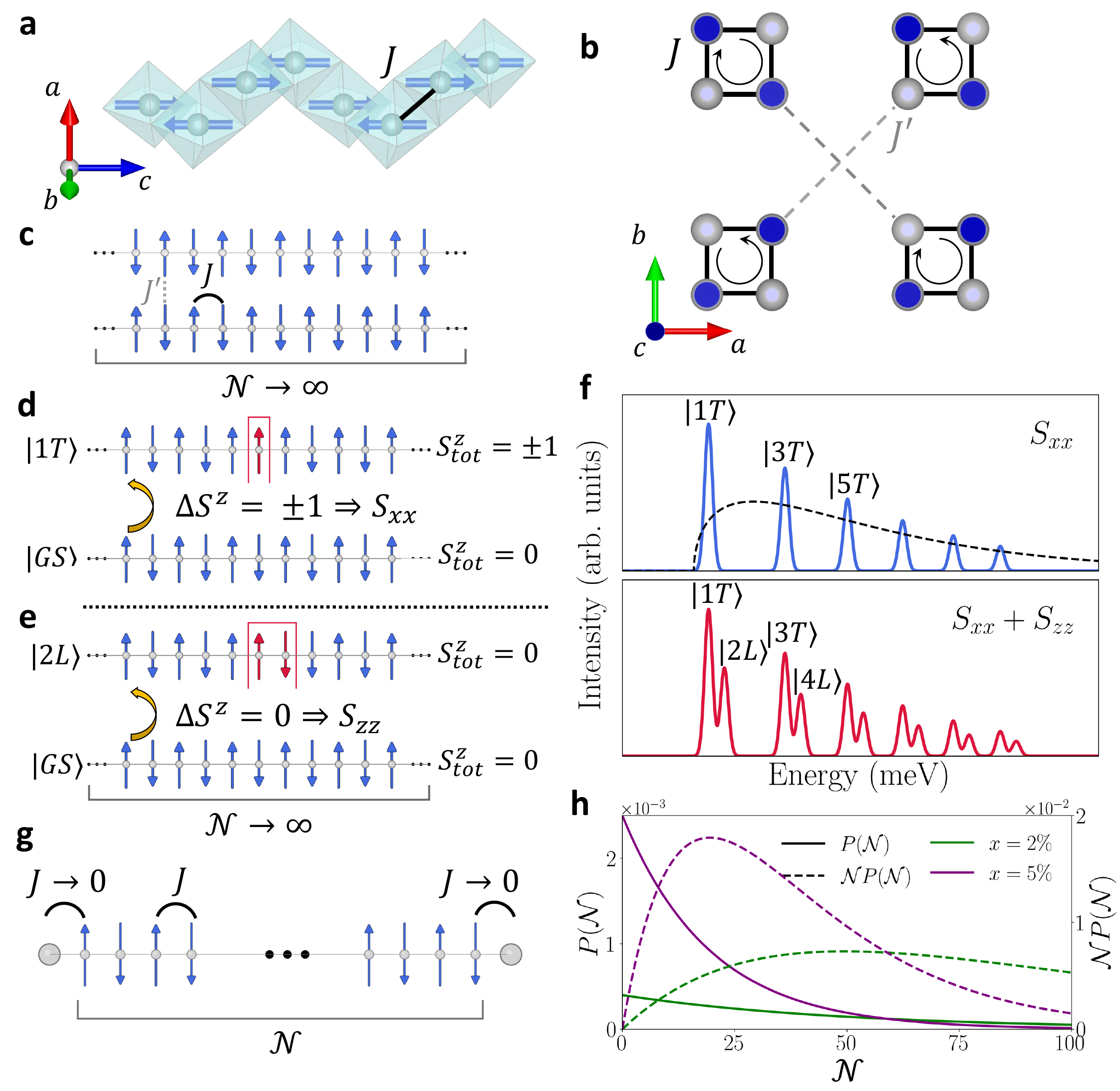}
\caption{\textbf{Magnetic model for \bacovo\ and its Mg-doped variant.}
(a) One-dimensional screw chain of edge-sharing Co$^{2+}$ octahedra along the crystallographic $c$-axis in \bacovo.
(b) Projection onto the $(a,b)$ plane, showing neighboring chains; the complex interchain exchange paths yield the average effective antiferromagnetic diagonal coupling $J'$ (dashed lines).
(c) Minimal magnetic model consisting of two antiferromagnetic $S=\tfrac{1}{2}$ chains with intrachain coupling $J$ and effective interchain coupling $J'$ (spins rotated by $90^{\circ}$ for clarity).
(d,e) Lowest confined excitations: the first transverse mode $\ket{1T}$ and the first longitudinal mode $\ket{2L}$ with $\Delta S^{z}=\pm 1$ and $0$ respectively with respect to the ground state.
(f) Schematic representation of Zeeman-ladder spectra of confined spinons, showing the transverse response ${\cal S}_{xx}$ (top) and the combined response ${\cal S}_{xx}+{\cal S}_{zz}$ (bottom). The dashed line corresponds to the gapped spinon-continuum expected for an Ising spin chain in absence of interchain coupling. 
(g) Mg substitution segments the chains into finite lengths $\cal{N}$, suppressing exchange near the boundaries ($J\rightarrow0$) and realizing open boundary conditions. 
(h) Geometrical distribution of segment lengths for diluted spin chains. Plain and dashed lines denote $P(\cal{N})$ and $\cal{N} P(\cal{N})$ respectively while green and purple colors denote $x = 2\%$ and $x = 5\%$ respectively.  
}
\label{Fig0}
\end{figure*}

We now consider the effects of doping by substituting magnetic Co$^{2+}$ ions with 
non-magnetic Mg$^{2+}$ ions. We assume that this substitution fragments the spin chains into finite segments of various lengths [Fig.~\ref{Fig0}g], each described by a Hamiltonian with 
open boundary conditions:
\begin{align}
 \mathcal{H}_{\mathcal{N}} =
  J \sum_{j=1}^{\mathcal{N}-1} \left[\epsilon
   \left(S_{j}^{x}S_{j+1}^{x}+S_{j}^{y}S_{j+1}^{y}\right)
   +S_{j}^{z} S_{j+1}^{z}\right]
   +h_{\mathrm{MF}} \sum_{j=1}^{\mathcal{N}}(-1)^{j}S_{j}^{z},
\label{eq:HamilN}
\end{align}
with $h_{\mathrm{MF}}=J'|\braket{S_{j}^{z}}|$ the effective staggered magnetic field 
from neighboring chains. The probability that a chain segment has a length $\mathcal{N}$, given that $x$ is the Mg$^{2+}$ concentration, is obtained by the geometric distribution~\cite{Eggert2002,Sirker2007}, $P(\mathcal{N}) = x(1-x)^{\mathcal{N}}$, 
with mean $\bar{\mathcal{N}} \approx 1/x$ in the dilute limit [Fig.~\ref{Fig0}h].

To experimentally realize such finite-size chain segments, \bacovomg\ single crystals 
at $x = 2\%$ ($\Leftrightarrow \mathcal{\bar{N}} = 50$) and $5\%$ ($\Leftrightarrow \mathcal{\bar{N}} = 20$) were synthesized via the optical floating-zone 
technique~\cite{Lejay2011} and characterized by neutron diffraction (see Supplementary 
Material). Both the crystallographic structure and the zero-field antiferromagnetic 
order remain essentially intact upon doping, confirming that Mg$^{2+}$ substitution 
cleanly realizes open boundary conditions without structural distortion. The N\'eel 
temperature and ordered magnetic moment decrease linearly with $x$, from 
$T_{\mathrm{N}} \simeq 5.5$~K and $2.2~\mu_{\mathrm{B}}$ in the parent compound to 
$T_{\mathrm{N}} = 3.8$~K and $1.3~\mu_{\mathrm{B}}$ at $x = 5\%$, consistent with 
previous macroscopic characterizations~\cite{Niesen2014}.

\subsection*{Emergence of localized excitations revealed by inelastic neutron scattering}
\label{sec:Experiments}

\begin{figure*}[thb]
\centering
\includegraphics[width=\linewidth]{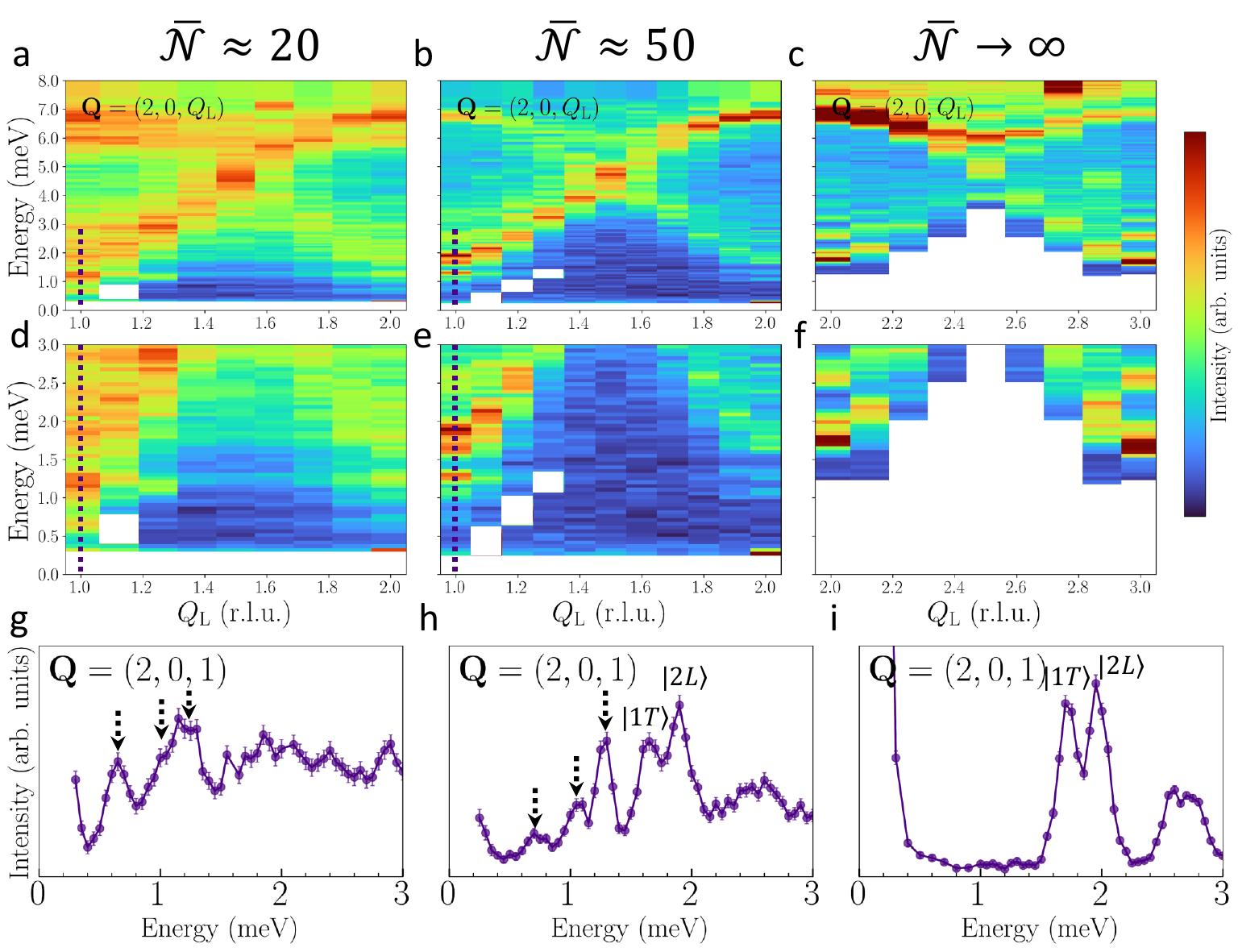}
\caption{\textbf{Spin dynamics of \bacovomg.} (a--c) Inelastic neutron scattering intensity maps of the intrachain dispersion measured along $\boldsymbol{Q}=(2,0,Q_{\rm L})$ at $T=1.5$~K for $x=5\%$, $2\%$, and $0$, corresponding to chains with average lengths $\bar{\mathcal{N}}\simeq20$, $50$, and $\infty$, respectively. Although the Brillouin zones displayed for the doped and undoped compounds differ, they are crystallographically equivalent. For clarity, the spurious (parasitic) intensity previously reported in Ref.~\cite{Faure2021_transverse}, extending from $\mathbf{Q}\approx(2,0,1.1)$ at $E\approx0.5$~meV to $\mathbf{Q}\approx(2,0,1.25)$ at $E\approx1.1$~meV, has been removed (see the Supplemental Material for details). (d--f) Enlarged views of panels (a--c) in the low-energy range between 0 and 3~meV. (g--i) Representative constant-$\boldsymbol{Q}$ energy scans used to construct the dispersion maps (taken at the vertical dashed lines in panels (a,b)). In the doped samples, additional nearly non-dispersive modes emerge at approximately $0.7$, $1.0$, and $1.2$~meV (indicated by the dashed black arrows in panels (g,h)), whereas the undoped compound exhibits only the Zeeman ladder excitations. Owing to the absence of magnetic intensity below the first transverse excitation $\ket{1T}$ observed in (i), the full low-energy dispersion was not measured for the undoped sample in panel (c).}
\label{Fig1}
\end{figure*}

Figures~\ref{Fig1}(a--b) present the INS intensity maps along the intrachain direction $\boldsymbol{Q} = (2, 0, Q_{L})$ for $x = 5\%$ ($\bar{\mathcal{N}}=20$) and $2\%$ ($\bar{\mathcal{N}}=50$) at $T = 1.5$~K. For comparison, the undoped compound [Fig.~\ref{Fig1}(c)] displays the Zeeman ladder spectrum with a spin gap $\Delta \simeq 1.75$~meV at the antiferromagnetic (AF) point $\boldsymbol{Q} = (2, 0, 1)$~\cite{Grenier2015}. In the doped samples, the Zeeman ladder modes persist: magnetic modes corresponding to the first transverse and longitudinal excitations are indeed observed around 1.7 and 1.9~meV for $x = 2\%$, and merged in a broad excitation around 1.6 to 1.8~meV for $x = 5\%$ at the AF point [Figs.~\ref{Fig1}(d--e). They disperse up to a maximum energy $E_m \simeq 4$~meV at $\boldsymbol{Q} = (2, 0, 1.5)$.
More strikingly, three additional features emerge \emph{below} the first Zeeman ladder at around 0.7, 1.0, and 1.2~meV [Figs.~\ref{Fig1}(d--e)], with intensity increasing with energy and fading away from AF and zone-center points. The absence of dispersion indicates a localized origin.

\begin{figure*}[thb]
\centering
\includegraphics[width=\linewidth]{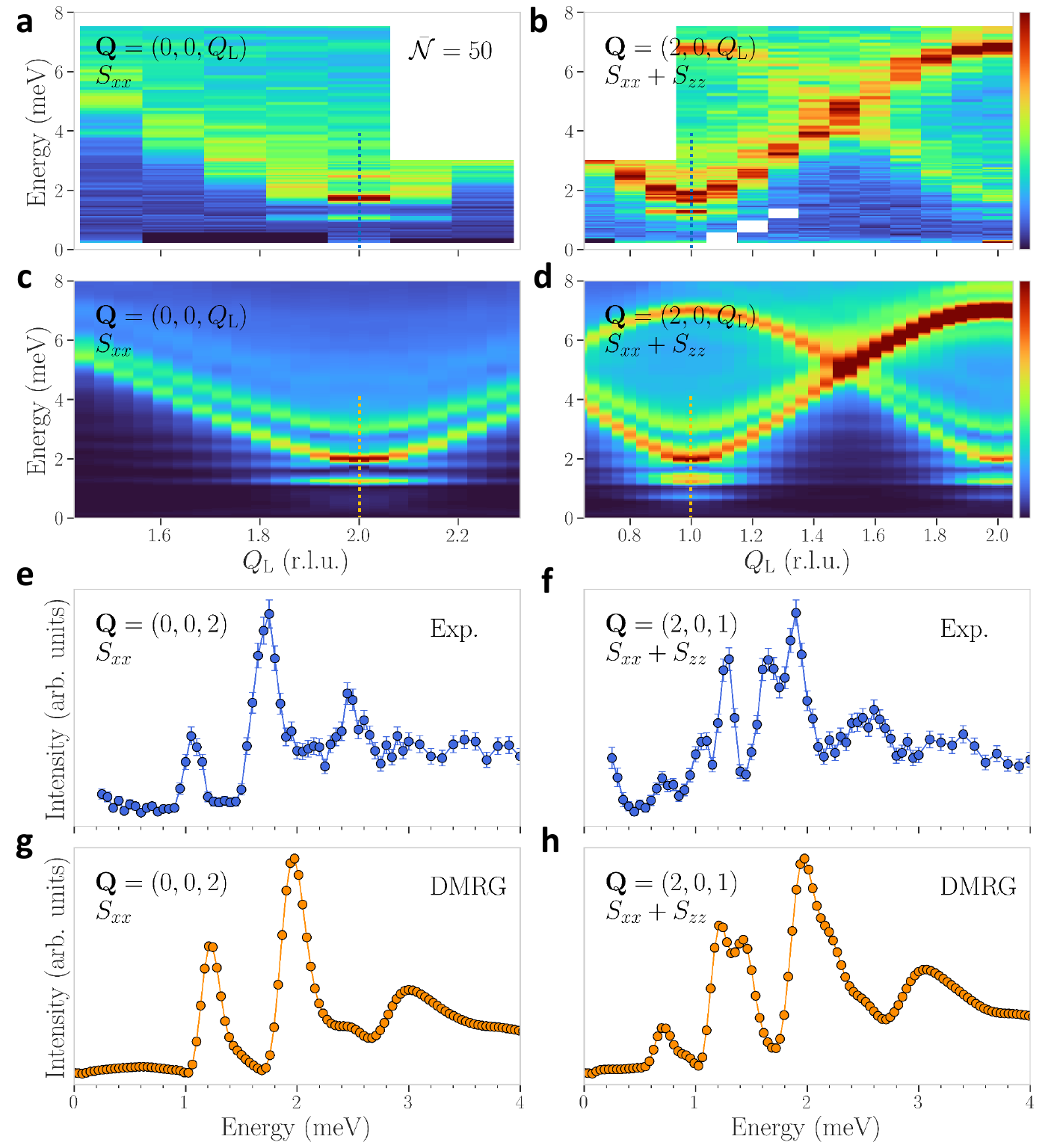}
\caption{\textbf{Transverse and longitudinal character of the emergent modes}: (a-b) Inelastic neutron scattering intensity maps of the intrachain dispersion along $\boldsymbol{Q} = (0, 0, Q_L)$ and $\boldsymbol{Q} = (2, 0, Q_L)$ for $x = 2\%$ ($\bar{\mathcal{N}} \simeq 50$) and measured at $T=1.5$~K. The former is sensitive to transverse fluctuations ${\cal S}_{xx}$ only while the later is related to both transverse ${\cal S}_{xx}$ and longitudinal ${\cal S}_{zz}$ fluctuations. (c-d) Numerically calculated DMRG (see text) corresponding dispersion maps. (e-h) Comparison between experimental (blue points) and calculated (orange points) intensity for $\mathbf{Q}$-constant energy scans at $\boldsymbol{Q} = (0, 0, 2)$ (e,g) and $\boldsymbol{Q} = (2, 0, 1)$ (f,h) (corresponding to the vertical dashed lines in panels (a--d)).}
\label{Fig2}
\end{figure*}

To characterize the transverse and longitudinal nature of these excitations, we compare INS spectra along $\boldsymbol{Q} = (0, 0, Q_{\rm L})$, sensitive to ${\cal S}_{xx}$ only, and $\boldsymbol{Q} = (2, 0, Q_{\rm L})$, probing both ${\cal S}_{xx}$ and ${\cal S}_{zz}$, with density matrix renormalization group (DMRG) calculations [Figs.~\ref{Fig2}(a--d)]. The calculations successfully reproduce the experimental dispersion maps, including the low-energy non-dispersive features and their momentum dependence. A direct comparison of energy scans at $\boldsymbol{Q} = (0, 0, 2)$ and $\boldsymbol{Q} = (2, 0, 1)$ is shown in Figs.~\ref{Fig2}(e--h). The main spectral features are well reproduced: at $\boldsymbol{Q} = (0, 0, 2)$ [Fig.~\ref{Fig2}e], a localized transverse mode is calculated around 1.2~meV (compared to 1~meV in the experiment), below the first transverse Zeeman ladder at 1.9~meV (compared to 1.7~meV); at $\boldsymbol{Q} = (2, 0, 1)$ [Fig.~\ref{Fig2}g], two additional longitudinal modes appear at 0.7 (compared to 0.7~meV) and 1.2~meV (compared to 1.4~meV). While the overall agreement is excellent, some discrepancies remain in the mode energies, slightly shifted upward in the calculations, and in relative intensities. This may arise from (i) mean-field treatment of the interchain coupling, neglecting weak interchain dispersion~\cite{faure_topological_bacovo}; (ii) impurity effects accounted for only in the main chain, overestimating magnetic order due to interchain coupling; (iii) $J$ and $J'$ fixed from the undoped compound, although doping likely affects these parameters, as suggested by the reduction of the ordering temperature and first transverse Zeeman gap when increasing $x$ [Figs.~\ref{Fig1}(a--c) and Supplementary Material].

\subsection*{Cluster and edge excitations: even-odd parity and chain-length dependence}

\begin{figure*}[thb]
\centering
\includegraphics[width=\linewidth]{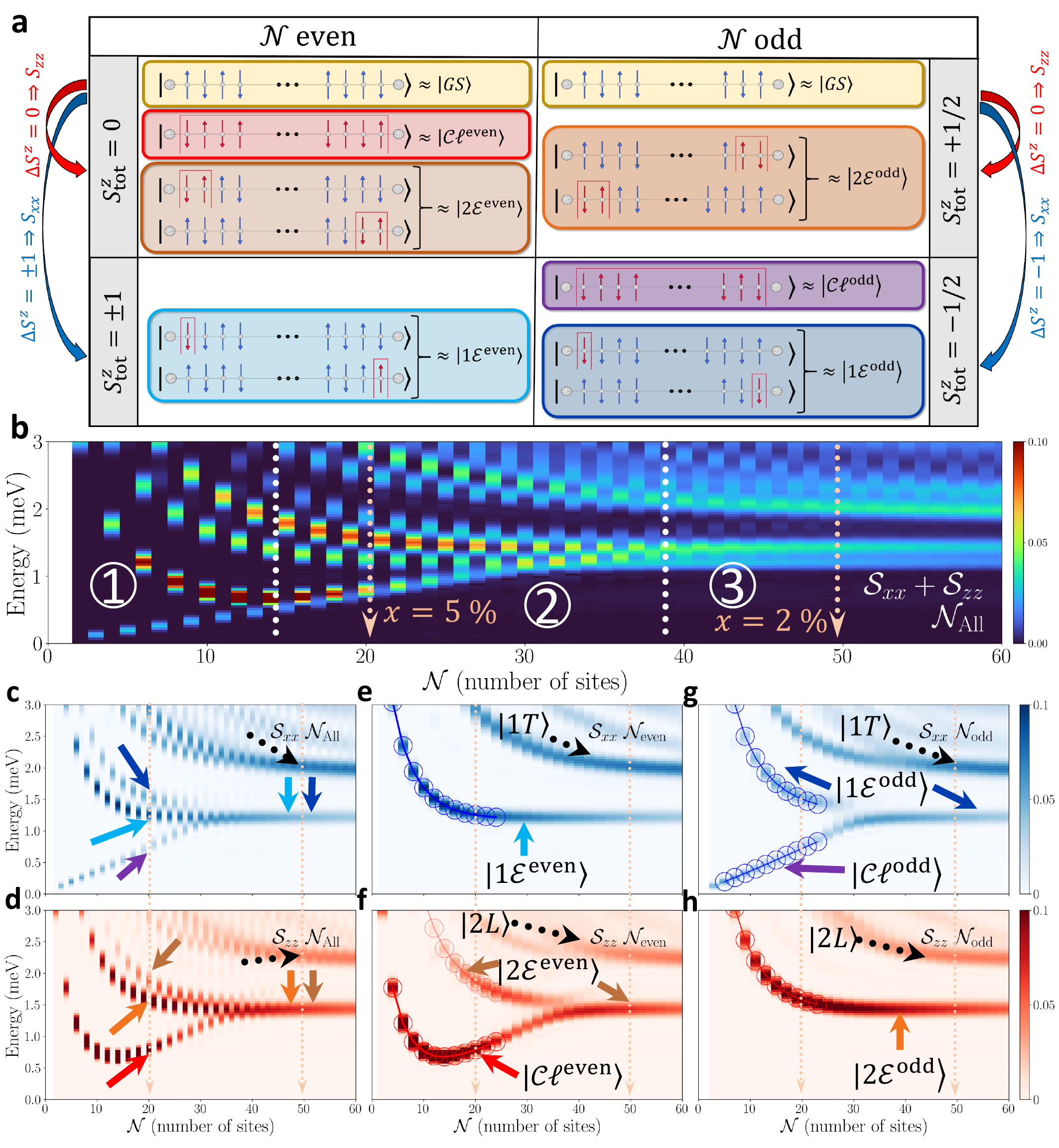}
\caption{
\textbf{Quantum nature of cluster and edge modes revealed by ED and DMRG calculations.} 
(a) Relevant Ising-limit basis configurations in the total-spin sectors $S^z_{\mathrm{tot}} = 0, \pm 1$ for even $\mathcal{N}$, and $S^z_{\mathrm{tot}} = \pm \tfrac{1}{2}$ for odd $\mathcal{N}$. These configurations provide a simple picture of the lowest-energy excitations, whose detailed weights are obtained from ED (see Supplementary Material). In regimes 2 and 3, the ground states $\ket{\mathrm{GS}}$ (for both even and odd chains) are dominated by one of the Néel configurations due to symmetry breaking induced by the interchain coupling (see text). Transitions with $\Delta S^z_{\rm tot}=\pm1$ ($\Delta S^z_{\rm tot}=0$) correspond to transverse (longitudinal) correlations, indicated by blue (red) arrows. (b) DMRG calculations of the summed transverse ${\cal S}_{xx}$ and longitudinal ${\cal S}_{zz}$ dynamical susceptibilities as a function of the total number of sites $\mathcal{N}$, evaluated at $\boldsymbol{Q} = (2, 0, 2)$. White dotted lines delimit regimes 1–3, and orange arrows mark the experimental impurity concentrations $x = 2\%$ and $x = 5\%$. (c–h) Transverse ${\cal S}_{xx}$ (upper panels) and longitudinal ${\cal S}_{zz}$ (lower panels) for: (c, d) all lengths, (e, f) even $\mathcal{N}$, and (g, h) odd $\mathcal{N}$. In (e–h), blue and red open circles denote ED results; lines are guides to the eye. Colored arrows indicate the excitations shown in (a).
}
\label{Fig3}
\end{figure*}

To investigate the impurity effects in detail, we performed exact diagonalization (ED) and DMRG calculations on finite-length chains with open boundary conditions. Figure~\ref{Fig3}b shows the numerically calculated sum of transverse ${\cal S}_{xx}$ and longitudinal ${\cal S}_{zz}$ dynamical susceptibilities as a function of the chain length $\mathcal{N}$, with the individual components for all, even, and odd $\mathcal{N}$ shown in Figs.~\ref{Fig3}(c--h). Low-energy modes are observed in the spectra. While the states depicted in Fig.~\ref{Fig3}a are well-defined configurations in the Ising limit, \bacovo\ lies away from this limit ($\epsilon=0.53$) and the transverse $XX$ exchange term mixes different spin configurations, as is also the case for the Zeeman ladders. Consequently, the actual eigenstates are linear combinations of several basis states. Here, for clarity, we restrict the discussion to the Ising-limit configurations shown in Fig.~\ref{Fig3}a, which provide a simple picture of the physics at play. The detailed weights of the different configurations and their evolution with $\mathcal{N}$ are presented in the Supplementary Material.

For $\mathcal{N}=\mathrm{even}$ chains, the two N\'{e}el-ordered configurations 
$\ket{\uparrow\downarrow\cdots\uparrow\downarrow}$ and $\ket{\downarrow\uparrow\cdots\downarrow\uparrow}$ both belong to the 
$S_{\mathrm{tot}}^{z}=0$ sector and are entangled by the $XY$ exchange term, which 
induces quantum fluctuations via successive two-spin flips $\uparrow\downarrow \;\Leftrightarrow\; \downarrow\uparrow$. This hybridization is maximal for short chains: for $\mathcal{N}=2$ and small interchain mean-field term $h_{\mathrm{MF}}$, it yields the antibonding singlet ground state $\ket{\mathrm{AB}_{\mathcal{N}=2}} \simeq (\ket{\uparrow\downarrow}-\ket{\downarrow\uparrow})/\sqrt{2}$ and the bonding triplet excited state $\ket{\mathrm{B}_{\mathcal{N}=2}} \simeq (\ket{\uparrow\downarrow}+\ket{\downarrow\uparrow})/\sqrt{2}$, both maximally entangled with negligible N\'{e}el order. As $\mathcal{N}$ increases, two effects progressively suppress the entanglement: the $XY$-induced fluctuations require an increasing number of successive two-spin flips to connect the two configurations, and the mean-field term $h_{\mathrm{MF}}$ increasingly splits their energies. As a result, the system displays a behavior reminiscent of Anderson's tower of excited states~\cite{Anderson1952, Sotnikov2023}: the quantum superpositions 
$\ket{\mathrm{AB}_{\mathcal{N}}}$ and $\ket{\mathrm{B}_{\mathcal{N}}}$ progressively 
evolve toward classical-like N\'{e}el states. The antibonding state $\ket{\mathrm{AB}_{\mathcal{N}}}$ asymptotically approaches the ground state 
$\ket{\uparrow\downarrow\cdots\uparrow\downarrow}$, stabilized by $h_{\mathrm{MF}}$, 
while the bonding state $\ket{\mathrm{B}_{\mathcal{N}}}$ asymptotically approaches the 
\emph{cluster} state $\ket{\mathcal{C\ell}^{\mathrm{even}}} \equiv \ket{\downarrow\uparrow\cdots\downarrow\uparrow}$, the reversed N\'{e}el configuration 
with respect to the ground state. Crucially, unlike the ideal Anderson tower scenario 
where the singlet-triplet gap closes in the thermodynamic limit, here $h_{\mathrm{MF}}$ 
explicitly breaks the symmetry between the two N\'{e}el configurations, so that the gap 
never closes. As seen in Fig.~\ref{Fig3}f, the gap decreases as $\sim 1/\mathcal{N}$ 
for $\mathcal{N} \lesssim 14$ (disentanglement regime) and reaches a minimum near 
$\mathcal{N} \sim 14$, then grows linearly as $\propto \mathcal{N} h_{\mathrm{MF}}$ for 
$14 \lesssim \mathcal{N} \lesssim 36$, reflecting the interchain mean-field energy cost 
of the transition from $\ket{GS}$ to $\ket{\mathcal{C\ell}^{\mathrm{even}}}$. Since 
this transition involves $\Delta S_{\mathrm{tot}}^z = 0$, it is visible in $\mathcal{S}_{zz}$.

For $\mathcal{N}=\mathrm{odd}$ chains, the two lowest N\'{e}el configurations 
$\ket{\uparrow\downarrow\cdots\uparrow\downarrow\uparrow}$ and 
$\ket{\downarrow\uparrow\cdots\downarrow\uparrow\downarrow}$ belong to different 
$S_{\mathrm{tot}}^z$ sectors ($+1/2$ and $-1/2$ respectively) and cannot be connected 
by any number of two-spin flips, so that they do not entangle. They correspond to the 
ground state $\ket{GS}$ and the cluster state $\ket{\mathcal{C\ell}^{\mathrm{odd}}}$, 
already well-defined even for the shortest chains, with a gap growing linearly with 
$\mathcal{N}$ throughout $\mathcal{N} \lesssim 36$ [Fig.~\ref{Fig3}g]. The transition 
between these states yields $\Delta S_{\mathrm{tot}}^z = \pm 1$ and appears in 
$\mathcal{S}_{xx}$.

Edge states arise because locating one domain wall at the chain boundary suppresses the 
energy cost of a spin-flip excitation, placing them below the Zeeman ladder modes. For 
$\mathcal{N}=\mathrm{even}$, the lowest edge mode $\ket{1\mathcal{E}^{\mathrm{even}}}$ 
($\Delta S_{\mathrm{tot}}^z = \pm 1$, visible in $\mathcal{S}_{xx}$) is generated by a 
single spin flip from the ground state; its energy decreases with system size up to 
$\mathcal{N} \sim 24$ and then converges to $\approx 1.2$~meV [Fig.~\ref{Fig3}e]. A 
second edge mode $\ket{2\mathcal{E}^{\mathrm{even}}}$ involves a two-spin-flip 
configuration ($\Delta S_{\mathrm{tot}}^z = 0$, visible in $\mathcal{S}_{zz}$) and 
carries the second longitudinal spectral weight [Fig.~\ref{Fig3}f]. Its energy decreases monotonically with 
$\mathcal{N}$ until an anticrossing with $\ket{\mathcal{C\ell}^{\mathrm{even}}}$ occurs 
near $\mathcal{N} \simeq 36$, marking the point at which a single edge spin flip becomes 
energetically favorable over reversing the entire chain, i.e. when $\mathcal{N}J' \simeq 
J$, giving $\mathcal{N} \simeq 34$ for $J=5.8$~meV and $J'=0.17$~meV, consistent with 
the numerical result. For longer chains, its energy converges to $\approx 1.4$~meV. The odd-chain case mirrors this picture: 
$\ket{1\mathcal{E}^{\mathrm{odd}}}$ ($\Delta S_{\mathrm{tot}}^z = \pm 1$, visible in 
$\mathcal{S}_{xx}$) undergoes an anticrossing with $\ket{\mathcal{C\ell}^{\mathrm{odd}}}$ 
near $\mathcal{N} \simeq 36$ and converges to $\approx 1.2$~meV [Fig.~\ref{Fig3}g], 
while the two-spin-flip edge mode $\ket{2\mathcal{E}^{\mathrm{odd}}}$ ($\Delta 
S_{\mathrm{tot}}^z = 0$, visible in $\mathcal{S}_{zz}$) shows no anticrossing and 
converges directly to $\approx 1.4$~meV for $\mathcal{N} \gtrsim 24$ [Fig.~\ref{Fig3}h].

From this analysis, three distinct regimes of chain-length dependence emerge, delimited 
by the white dotted lines in Fig.~\ref{Fig3}b. \textbf{(1) Short-chain regime} 
($\mathcal{N} \lesssim 14$): the spectrum is strongly parity-dependent; for even 
$\mathcal{N}$, bonding and antibonding states dominate and progressively disentangle, 
while for odd $\mathcal{N}$, the N\'{e}el and cluster states already constitute the 
dominant contributions to the ground state and first excited state, respectively, even 
for the shortest chain lengths. \textbf{(2) Intermediate regime} 
($14 \lesssim \mathcal{N} \lesssim 36$): well-defined cluster and edge states coexist 
with length-dependent energy levels; even/odd distinctions persist but are reduced. 
\textbf{(3) Long-chain regime} ($\mathcal{N} \gtrsim 36$): the anticrossings have 
occurred, parity dependence vanishes, and the system approaches the thermodynamic limit 
with size-independent edge mode energies. In the infinite-chain limit, spectral weight 
transfers entirely to the Zeeman ladder modes $\ket{1T}$ and $\ket{2L}$ at $\approx 2.0$ 
and $2.2$~meV, and edge-mode weight vanishes.

\subsection*{Neutron spectroscopic identification of finite-chain excitations}

\begin{figure*}[thb]
\centering
\includegraphics[width=1\linewidth]{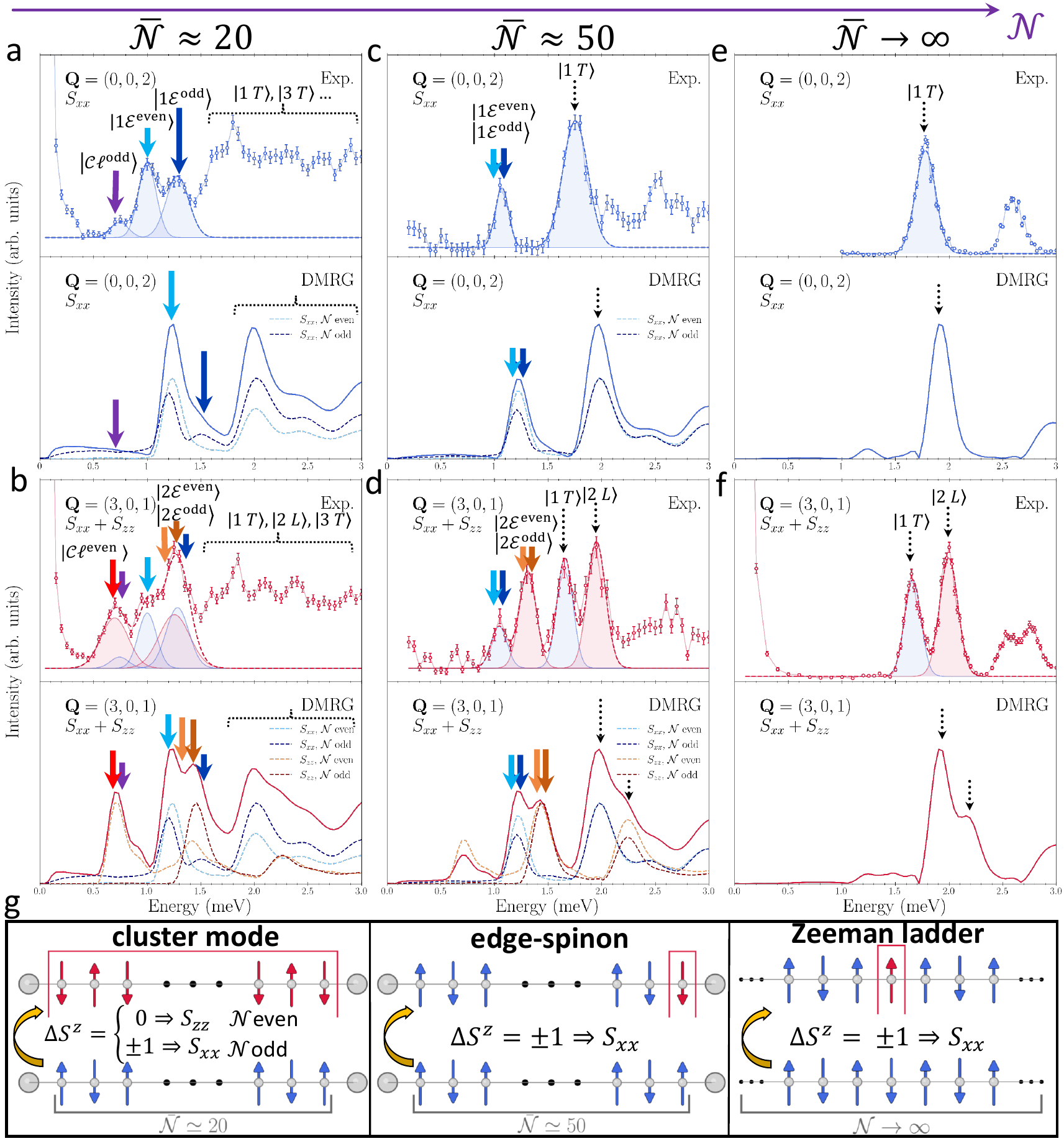}
\caption{\textbf{Evolution of the lowest-energy modes with increasing chain length.}
(a--f) INS spectra (open circles) and DMRG calculations (thick solid lines) for $x=5\%$, $2\%$, and $0$, corresponding to average chain lengths $\bar{\mathcal{N}}\approx20$, $50$, and $\infty$, respectively. Thin solid lines in the experimental (Exp.) panels show fits to the INS data, highlighting the excitation modes identified by DMRG. Dashed lines below the DMRG spectra show the even- and odd-chain contributions to ${\cal S}_{xx}$ (light and dark blue) and ${\cal S}_{zz}$ (orange and brown), respectively, with intensities multiplied by a factor of 3 for clarity. Measurements and calculations at $\mathbf{Q}=(0,0,2)$ (blue) and $\mathbf{Q}=(3,0,1)$ (red) probe ${\cal S}_{xx}$ and ${\cal S}_{xx}+{\cal S}_{zz}$, respectively. Colored arrows indicate the excitations identified by ED and DMRG, following the color code of Fig.~\ref{Fig3}; black arrows mark the Zeeman ladders. (g) Schematic evolution of the lowest-energy excitations with chain length. At $x=5\%$ (regime 2), the cluster mode $\ket{\mathcal{C \ell}^{\rm odd}}$ and $\ket{\mathcal{C \ell}^{\rm even}}$, with transverse or longitudinal character depending on chain parity. At $x=2\%$ (regime 3), bound edge spinons $\ket{1\mathcal{E}^{\rm even}}$ and $\ket{1\mathcal{E}^{\rm odd}}$ (now indistinguishable) , corresponding to a single spin flip at the chain boundary and therefore a transverse transition ($\Delta S^z=\pm1$). In the infinite-chain limit ($x = 0$), the first transverse Zeeman ladder, $\ket{1T}$ ($\Delta S^z=\pm1$).}
\label{Fig4}
\end{figure*}

Based on the above analysis of our numerical work, which reveals the evolution of the excitation spectrum, from a quantum cluster mode to edge-localized states, we can now interpret the results of INS experiments.
In actual doped materials, the experimental dynamical susceptibility naturally includes contributions from segments of different lengths. The measurements performed for $x = 5\%$ and $2\%$ thus probe two different chain-length regimes. This summation is also taken into account in calculations, each individual chain segment being weighted by the geometric distribution of lengths (and dominated by segments of length $\bar{\mathcal{N}} \approx 1/x$). A direct comparison between theory and experiment makes it possible to identify the nature of the observed excitations [Fig.~\ref{Fig4}].

For $x=5\%$, corresponding to $\bar{\mathcal{N}}\simeq 20)$ [Fig.~\ref{Fig4}(a,b)], the system lies in the \textbf{intermediate chain regime}, where the excitation spectrum retains a clear chain-parity dependence. The ${\cal S}_{xx}$ component probed at $\mathbf{Q}=(0,0,2)$ [Fig.~\ref{Fig4}a] reveals three transverse excitations at approximately 0.7, 1.0 and 1.25~meV, assigned to $\ket{\mathcal{C \ell}^{\rm odd}}$, $\ket{1\mathcal{E}^{\rm even}}$, and $\ket{1\mathcal{E}^{\rm odd}}$ respectively, based on calculations for ideal chains of $ \mathcal{N}=20$ and $\mathcal{N} = 21$ sites [Fig.~\ref{Fig3}c]. Notably, both the well-defined cluster-mode feature and the strong intensity of the 1.25~meV excitation exceed expectations from calculations averaging over a geometrical chain-length distribution [Fig.~\ref{Fig4}a]. This may indicate a higher proportion of chains around the mean length of 20 sites, possibly related to the small difference in ionic radii between Mg$^{2+}$ and Co$^{2+}$, although this interpretation remains speculative. At $\mathbf{Q}=(3,0,1)$ [Fig.~\ref{Fig4}(b)], both ${\cal S}_{xx}$ and ${\cal S}_{zz}$ contribute. The most prominent additional feature is the even-chain cluster mode $\ket{\mathcal{C \ell}^{\rm even}}$ at around 0.7~meV, which carries significantly larger spectral weight than the odd-chain cluster mode, consistent with the DMRG results of Fig.~\ref{Fig4}b. Additional longitudinal excitations associated with $\ket{2\mathcal{E}^{\rm even}}$ and $\ket{2\mathcal{E}^{\rm odd}}$ are observed around 1.25~meV. Although their parity-dependent energy splitting is difficult to resolve, an additional broader Gaussian component is required to reproduce the data, reflecting the distribution of short chain segments and the associated energy broadening. Above 1.5~meV, the response includes contributions from a distribution of transverse Zeeman ladders, also reproduced by the distribution-averaged calculations. Although a full separation of transverse and longitudinal contributions would require polarized neutron scattering, the overall mode assignment is supported by the DMRG calculations in Fig.~\ref{Fig4}b. Altogether, these observations provide a clear experimental signature of the parity-dependent excitation spectrum characteristic of the \textbf{intermediate chain regime} as well as unambiguously establishing the presence of a cluster mode.

For $x=2\%$, corresponding to $\bar{\mathcal{N}}\simeq50$ [Fig.~\ref{Fig4}(c,d)], the system lies in the \textbf{long chain regime} where the excitation spectrum becomes essentially independent of chain parity. At $\mathbf{Q}=(0,0,2)$, a transverse excitation is observed near 1.05~meV, which corresponds to $\ket{1\mathcal{E}^{\mathrm{even}}}$ and $\ket{1\mathcal{E}^{\rm odd}}$, now indistinguishable as the system approaches the thermodynamic limit. At $\mathbf{Q}=(3,0,1)$, an additional measured excitation around 1.25~meV matches the now-degenerate $\ket{2\mathcal{E}^{\rm even}}$ and $\ket{2\mathcal{E}^{\rm odd}}$ modes. A residual peak around 0.7~meV, visible in both channels and more clearly in Figs.~\ref{Fig2}f and \ref{Fig2}h, is the remnant cluster-mode signal from short segments, which still exist with small probability in the geometric distribution. The second lowest excitation around 1.7~meV corresponds to the first Zeeman ladder state $\ket{1T}$. These observations confirm that the parity dependence characteristic of the \textbf{intermediate chain regime} disappears in the \textbf{long chain regime}.

Finally, in the pure compound ($x=0$), corresponding to the infinite-chain limit, the in-gap excitations vanish and the spectrum recovers the Zeeman ladder structure [Fig.~\ref{Fig4}(e,f)].

\section*{Discussion}

Previous theoretical work predicted edge-mode physics in anisotropic open chains~\cite{Wessel2000,Haque2010,Kattel2025_2}, and discrete excitations were experimentally observed in Mg-doped quasi-one-dimensional Ising systems~\cite{Nagler1983,Itoh1994_CsCoCl3,Yang2002_CsCoBr3}. In this context, the key finding of our work is that by varying the Mg-doping concentration, it becomes possible to reveal three distinct regimes in which chain parity, spin entanglement, and the nature 
of the dominant excitations all vary: a cluster mode corresponding to coherent reversal of the N\'{e}el segment, and edge states corresponding to spinons localized at segment boundaries. The INS spectra provide direct evidence for these distinct modes associated 
with even and odd chain lengths.

The experimental spectra are qualitatively captured by $XXZ$ calculations with open 
boundary conditions and mean-field interchain coupling. Although approximate — 
neighboring chains are treated as infinite — this approach captures the essential 
low-energy physics. Future work could involve cluster mean-field or two-dimensional DMRG 
approaches. On the experimental side, higher dopings ($\sim 10$--$15\%$) would access 
the short-chain regime, where the two N\'{e}el configurations of an even-length chain 
become strongly entangled, potentially forming a mesoscopic Schr\"{o}dinger cat state. 
Substitution with magnetic impurities may further open a route to Kondo-like physics in 
segmented quantum magnets~\cite{Eggert2002,Laflorencie2008}.

At a fundamental level, our results demonstrate how finite size, anisotropy, and open 
boundaries conspire to generate emergent boundary-localized excitations. The cluster mode 
in particular underscores the essential role of interchain coupling, which acts as an 
effective staggered field. More broadly, our findings could motivate studies in engineered 
quantum platforms — ultracold atoms, Rydberg arrays, and trapped-ion 
chains~\cite{Simon2011,Bernien2017,Schauss_2018} — where finite-length spin systems with 
tunable boundaries and staggered fields could directly realize edge spinons and cluster 
modes. Related boundary fractionalization has recently been observed in atomically 
engineered nanographene Haldane chains~\cite{Mishra2021}. From a quantum-information 
perspective, edge spinon modes offer localized, boundary-confined excitations suited for 
robust quantum state storage, while the cluster mode provides a segment-wide collective 
excitation for encoding quantum states across a finite chain, complementing efforts toward 
protected qubits in one-dimensional systems~\cite{Monroe2021,King2022,Mi2024}.

Our observations also raise the question of whether these modes can be interpreted within 
the framework of Majorana fermions. Via a Jordan--Wigner transformation, the system maps 
onto an interacting Majorana chain with spatially varying masses (see Supplementary 
Material), and \bacovo\ has recently attracted attention as a platform for Kitaev-type 
anisotropic interactions~\cite{Morris2021,Bhullar2025,Tang2025_XXZ}. A detailed analysis 
of these connections is left for future work. Finally, doping-induced segmentation in 
higher-dimensional correlated materials may similarly generate confined low-energy 
excitations, offering new perspectives on emergent phenomena in complex quantum matter.

In summary, combining neutron spectroscopy with numerical calculations, we demonstrate 
that chemical segmentation of a quasi-one-dimensional Ising antiferromagnet fundamentally 
reshapes its excitation spectrum through controlled finite-size effects. We identify three 
regimes governed by chain length: a short-chain regime characterized by strong 
entanglement and Anderson-tower physics, an intermediate regime dominated by a collective 
cluster mode, and a long-chain regime hosting edge-localized spinon modes. Our experiments 
directly probe the latter two, providing spectroscopic evidence for nondispersive 
excitations with clear chain-parity signatures, in close agreement with theory. These 
results establish segmented spin chains as a model platform for mesoscopic quantum 
magnetism and demonstrate that chemical substitution provides a controlled route to 
engineering collective and localized quantum degrees of freedom within a single bulk 
material.

\section*{Methods}

\noindent \textbf{Sample growth}

\bacovomg\ $(x = 2\%, 5\%)$ single-crystals were grown at Institut N\'{e}el by floating zone method. For both concentrations, a 5~cm long cylindrical crystal rod of about 4~mm diameter was obtained by imposing the growth axis to be along the ${\bf b}$ crystallographic axis. One crystal piece of 5 mm long was cut from the rod for the diffraction experiment while two crystal pieces of about 2 cm long were cut for the inelastic neutron scattering (INS) experiments.

\noindent \textbf{Neutron diffraction}

Two single-crystal diffraction experiments were carried out on the two thermal single crystal diffractometers 6T2 (LLB) for $x = 5\%$ and D23 (ILL) for $x = 2\%$. These experiments were performed with a 6 T and a 12 T cryomagnet respectively and with a dilution insert allowing to reach temperatures down to 50~mK. For 6T2 and D23, we used a Cu (200) monochromator with wavelength of 0.9~\AA and 1.28~\AA respectively.

\noindent \textbf{Inelastic neutron scattering experiment}

The INS experiments for both $x = 2\%$ and $x = 5\%$ were carried out on the cold-neutron triple-axis spectrometer FZJ-CRG IN12 at ILL. We used a combination of PG(002) monochromator and analyzer allowing to get an energy resolution of the order of 0.035~meV and 0.1~meV with final wavelength 1.05~\AA\ and 1.3~\AA\ respectively. The higher order contamination was suppressed using a velocity selector. The sample was oriented with the $\bm{b}$ axis vertical yielding a ($\bm{a}^*, \bm{c}^*$) horizontal scattering plane and was inserted in an orange cryostat allowing to reach a lowest temperature of 1.5~K.

\noindent \textbf{Inelastic neutron scattering formalism}

The spectra obtained in the INS experiments are related to the dynamical susceptibility, 
\begin{align}
 S_{\alpha\beta}(\boldsymbol{Q},\omega,\mathcal{N})
   =-\mathrm{Im}\int dt\sum_{j=1}^{\mathcal{N}}
   e^{i(\omega t-\boldsymbol{Q} \cdot \boldsymbol{r}_{j})}
   C_{\alpha\beta}^{\mathrm{R}}(\boldsymbol{r}_{j},t),
\end{align}
which is the Fourier transform of the retarded spin-spin correlation,
\begin{align}
 C_{\alpha\beta}^{\mathrm{R}}(\boldsymbol{r}_{j},t)
   =-i\vartheta (t)
     \braket{[S^{\alpha}(\boldsymbol{r}_{j},t),S^{\beta}(\boldsymbol{0},0)]},
\end{align}
where $\boldsymbol{r}_{j}$ is the position of Co$^{2+}$ ions, $\alpha, \beta = x, y,z$ and $\vartheta (t)$ is the step function. After taking the summation over the ensemble of chain segments, we obtain 
\begin{equation}
\mathcal{S}_{\alpha\beta}(Q,\omega) = \sum_{\mathcal{N}} \mathcal{N} P(\mathcal{N}) S_{\alpha\beta}(\boldsymbol{Q},\omega,\mathcal{N}),
\end{equation}
Note that the contribution from each segment is proportional to the length $\mathcal{N}$ of the segment. The graph of the weight $\mathcal{N} P(\mathcal{N})$ is given in Fig.~\ref{Fig0}h.

 At a given momentum transfer, the measured neutron instensity is proportional to the scattering function 
\begin{align}
 \mathcal{S}(\boldsymbol{Q},\omega)
 =\sum_{\alpha,\beta=x,y,z}
   \Big(\delta_{\alpha\beta}-\frac{Q_{\alpha}Q_{\beta}}{|\boldsymbol{Q}|^{2}}\Big)
   |f(\boldsymbol{Q})|^{2} \mathcal{S}_{\alpha\beta}(\boldsymbol{Q},\omega),
\label{eq:CrossSec}
\end{align} 
where $f(\boldsymbol{Q})$ is the Co$^{2+}$ magnetic form factor and $\boldsymbol{Q}$ the scattering vector. For $\boldsymbol{Q} \parallel$ $z$ axis, only the transverse component $\mathcal{S}_{xx}(\boldsymbol{Q},\omega)$ ($=\mathcal{S}_{yy}(\boldsymbol{Q},\omega)$) is captured. Otherwise, both longitudinal $\mathcal{S}_{zz}(\boldsymbol{Q},\omega)$ and transverse $\mathcal{S}_{xx}(\boldsymbol{Q},\omega)$ excitations are observed.

\noindent \textbf{Numerical calculations}

The inelastic neutron scattering spectra were calculated using the Chebyshev matrix product state (CheMPS) method, which combines density matrix renormalization group (DMRG) calculations for the ground state with a kernel polynomial expansion of the dynamical response. The many-body wavefunction was represented as a matrix product state. Ground states were obtained by finite-system DMRG on clusters of length $\mathcal{N}$ with open boundary conditions. Dynamical spin structure factors were then evaluated at zero temperature by applying the corresponding spin operator to the ground state and generating the Chebyshev vectors recursively after rescaling the Hamiltonian to the required interval. The spectra were reconstructed from the Chebyshev moments using a Jackson kernel. Exact diagonalization (ED) calculations were performed for chains of up to 24 sites to characterize the eigenstates in terms of their Ising-limit basis states. The principle of the numerical calculations and the used parameters are given in Supplementary Material. All calculations were performed using an in-house code developed by the authors.

\section*{Acknowledgments}

We acknowledge insightful discussions with Nicolas Laflorencie and Pierre Pfeuty. The neutron beam time was provided by the ILL and LLB neutron sources. ST is supported by a Grant-in-Aid for Scientific Research from JSPS, KAKENHI Grant Nos. JP24H00191, JP24K06891, and JP26K00662. This work was supported in part by the Swiss National Science Foundation under Division II (Grant No. 200020- 219400).

\section*{Author contributions}
All authors contributed significantly to this work. In details, sample preparation by PL, neutron scattering experiments and analysis by QF, BG, SP, and VS with the support of SR, ER and XF. Exact diagonalization calculations by ST, QF, SP and DMRG by ST. Physical discussions with TG; Manuscript written by QF with constant feedback from the other co-authors.

\section*{Competing interests}
The authors declare no competing financial interests.

\section*{Data availability}
Data are available upon request.

\bibliographystyle{naturemag}
\bibliography{bibliobis.bib}

\end{document}